\documentclass[aps,prl, superscriptaddress, 12pt]{revtex4}
\usepackage{xcolor}
\newif\ifproofread
\newcommand{\changemarker}[1]{%
	\ifproofread
	\textcolor{black}{#1}%
	\else
	#1%
	\fi}
\usepackage{graphicx}
\usepackage{amsmath}
\usepackage{amssymb}
\usepackage{colordvi}
\usepackage{mathrsfs}
\usepackage{bm}
\usepackage{bm}

\usepackage{epsfig}
\usepackage{natbib}
\usepackage{bibentry}
\begin{document}
\title{Design of graphene waveguide: Effect of edge orientation and waveguide configuration}
\author{Nayyar Abbas Shah}
\affiliation{CAS Key Laboratory of Quantum Information, and Synergetic Innovation Center of Quantum Information and Quantum Physics, University of Science and Technology of China,Chinese Academy of Sciences, Hefei 230026, China}
\author{Vahid Mosallanejad}
\email[Correspondence author:~~]{vahid@ustc.edu.cn}
\affiliation{CAS Key Laboratory of Quantum Information, and Synergetic Innovation Center of Quantum Information and Quantum Physics, University of Science and Technology of China,Chinese Academy of Sciences, Hefei 230026, China}
\author{Kuei-Lin Chiu}
\email[Correspondence author:~~]{eins0728@gmail.com}
\affiliation{Shenzhen Institute for Quantum Science and Engineering, Southern University of Science and Technology, Shenzhen 518055, China}
\author{Guo-ping Guo}
\email[Correspondence author:~~]{gpguo@ustc.edu.cn}
\affiliation{CAS Key Laboratory of Quantum Information, and Synergetic Innovation Center of Quantum Information and Quantum Physics, University of Science and Technology of China,Chinese Academy of Sciences, Hefei 230026, China}
\date{\today}
\begin{abstract}
Electron transport in a graphene quantum well can be \proofreadtrue\changemarker{analogous} to photon transmission in \proofreadtrue\changemarker{an} optical fiber. \proofreadtrue\changemarker{In this work, we present a detailed theoretical analysis to study the transport characteristics of graphene waveguides under the influence of different edge orientations. Non-equilibrium Green's function approach in combination with tight-binding Hamiltonian has been utilized to investigate the conductance properties of straight armchair and zigzag oriented graphene waveguides. Conductance plateaus at integer steps of $4e^2/h$ have been observed in  both orientations while the zigzag oriented waveguides present a wider first quantized plateau compared to that in the armchair oriented ones. Using various geometric and physical parameters, including side-barrier and waveguide width, and the metallic properties of terminals, we investigate the conductance profile of waveguides. In addition to the observation of valley-symmetry in both edge orientations, this article explores the critical influence of drain contacts on waveguide conductance. Furthermore, we extended our transport study to three different highly bent waveguide configurations, such as U-shape, L-shape and split-shape waveguides, in order to explore their applications in graphene-based ballistic integrated circuit devices.} \changemarker{In the end, we also calculated the conductance of larger graphene waveguides using the scalable tight-binding model, in order to compare the results obtained from the original model.} 
\end{abstract}
\maketitle

\section[sec:level1]{I. INTRODUCTION}

\proofreadtrue\changemarker{Ballistic transport and coherent conductance quantization are the key elements for engineering sophisticated nanoelectronic devices in new classes of materials ~\cite{van1988quantized, wharam1988one,somanchi2017diffusive, akbar2015graphene,connolly2013gigahertz,zhang2014measuring,chiu2017single}.}
Physically tailored graphene channels with \proofreadtrue\changemarker{widths} less than 50~nm, often noted as graphene nanoribbons (GNRs), provide an opportunity to \proofreadtrue\changemarker{manipulate} the electrical properties of the intrinsically gapless crystal~\cite{fujita1996peculiar,son2006energy,duerr2012edge,atteia2017ballistic}.
Electronic properties and stability of GNRs have been investigated for realistic applications such as transistors, filters and polarizers~\cite{barone2006electronic,schwierz2010graphene,ruffieux2012electronic, bao2011broadband,zhu2014electrically}.
\proofreadtrue\changemarker{The two well-known edge configurations, i.e., armchair and zigzag, result in two distinct forms of GNRs (commonly abbreviated by AGNRs and ZGNRs)~\cite{han2007energy, hancock2010generalized}.}
Transport properties in these two structures are different in many aspects, such as the spacing between conductance plateaus.
\proofreadtrue\changemarker{Although ideal GNRs should possess the quantization of conductance, unavoidable disorders on the edges have become dominant sources of incoherent scattering, making the quantization of conductance hardly visible in plasma-etched GNRs~\cite{li2008quantum,evaldsson2008edge,djavid2014computational,han2010electron, baldwin2016effect}}. \proofreadtrue\changemarker{To date, only few investigations into conductance quantization in GNRs fabricated using shadow mask oxygen plasma etching exist~\cite{lian2010quantum,lin2008electrical}}. Further improvement is now incorporated into the design of graphene point contacts and GNRs by using hexagonal-born-nitride \proofreadtrue\changemarker{as bottom and top dielectrics to reduce substrate disorders }~\cite{goossens2012gate,terres2016size,overweg2017electrostatically}. However, the pronounced quantization of conductance (mostly \proofreadtrue\changemarker{appearing as} kinks) is not easily accessible due to the hypersensitivity of the system to edge disorders~\cite{tombros2011quantized,clerico2018quantized}. 
\proofreadtrue\changemarker{On the other hand, charge carriers in graphene revealed phenomena such as refraction, reflection and Fabry-P\'erot interference that can be analogous to electromagnetic phenomena~\cite{rickhaus2015gate,chen2016electron,rickhaus2013ballistic}.} 
\proofreadtrue\changemarker{It has also recently been shown that the long phase coherence length in graphene embedded in van der Waals heterostructures provides unique opportunities to observe electron interference and other peculiar electron transmission states such as the snake states ~\cite{wei2017mach,makk2018coexistence,zebrowski2018aharonov}.}
\proofreadtrue\changemarker{The optics-like phenomena of electrons in graphene enables the design of all graphene electronic devices resembling an optical fiber, which effectively works as an electron waveguide ~\cite{cheianov2007focusing,low2009electronic,wu2011electronic}.}
\proofreadtrue\changemarker{When a uniform potential well is imposed across a graphene flake, the induced 1D quantum confinement in 2D electron gas results in straight graphene waveguides which have been explored both theoretically and experimentally with middle-scale (sub-micron size) and large-scale (micron size) geometries~\cite{petrovic2015fano, hartmann2010smooth,williams2011gate,rickhaus2015guiding,he2015guided,liu2015scalable}.} 
\proofreadtrue\changemarker{In line with the aforementioned theoretical studies, we have previously demonstrated that the quantization of conductance can be achieved in straight and bent armchair graphene waveguides by using Non-equilibrium Green's function (NEGF) calculation and proper design of contacts~\cite{kim2016valley,cao2017investigation,mosallanejad2018perfectly}.}
\proofreadtrue\changemarker{Recent work in the field studies also suggests that the connection between the external electrodes and the ribbon scattering area plays an important role in the conductance of GNRs~\cite{verges2018conductance, stegmann2017transport}.} 
\proofreadtrue\changemarker{Since AGNRs and ZGNRs have very different transport properties, we aim to address the question: what are the differences in transport between armchair-oriented and zigzag-oriented graphene waveguides (abbreviated as AO-GWs and ZO-GWs, respectively) with similar sizes?}
\proofreadtrue\changemarker{Our study includes two main parts. Firstly, we present a theoretical comparison between transport in straight AO-GWs and ZO-GWs. Secondly, we investigate the transmission characteristics of graphene waveguides with different geometries (L-shape, U-shape and split-shape), which had been previously studied in tailored graphene systems~\cite{xie2009effect,da2015curvature}.}
\proofreadtrue\changemarker{We organize this article in the following way: the geometry of AO-GW and ZO-GW and the details of our model are presented in section II.}
\proofreadtrue\changemarker{Conductance and local density of state are compared for straight AO-GW and ZO-GW in the first part of section III, where the corresponding quasi-one dimensional band structures for slices of waveguides are calculated for reference.}
\proofreadtrue\changemarker{Furthermore, the effect} of geometrical parameters such as the \proofreadtrue\changemarker{widths of side-barriers}, \proofreadtrue\changemarker{waveguide (potential well)} and terminals were investigated. \proofreadtrue\changemarker{Similar transport studies were also} carried out for L-shape, U-shape and split \proofreadtrue\changemarker{graphene waveguides}. \proofreadtrue\changemarker{The results are presented in the second part of section III.} \proofreadtrue\changemarker{In addition, the scalable tight-binding method has been utilized to examine the quantization of conductance for larger graphene waveguides in the last part of section III}. Finally, we will provide conclusive remarks about all waveguide configurations in section IV.
\section{II. Device Description and Methodology}

\begin{figure}
\includegraphics[width=8.6cm]{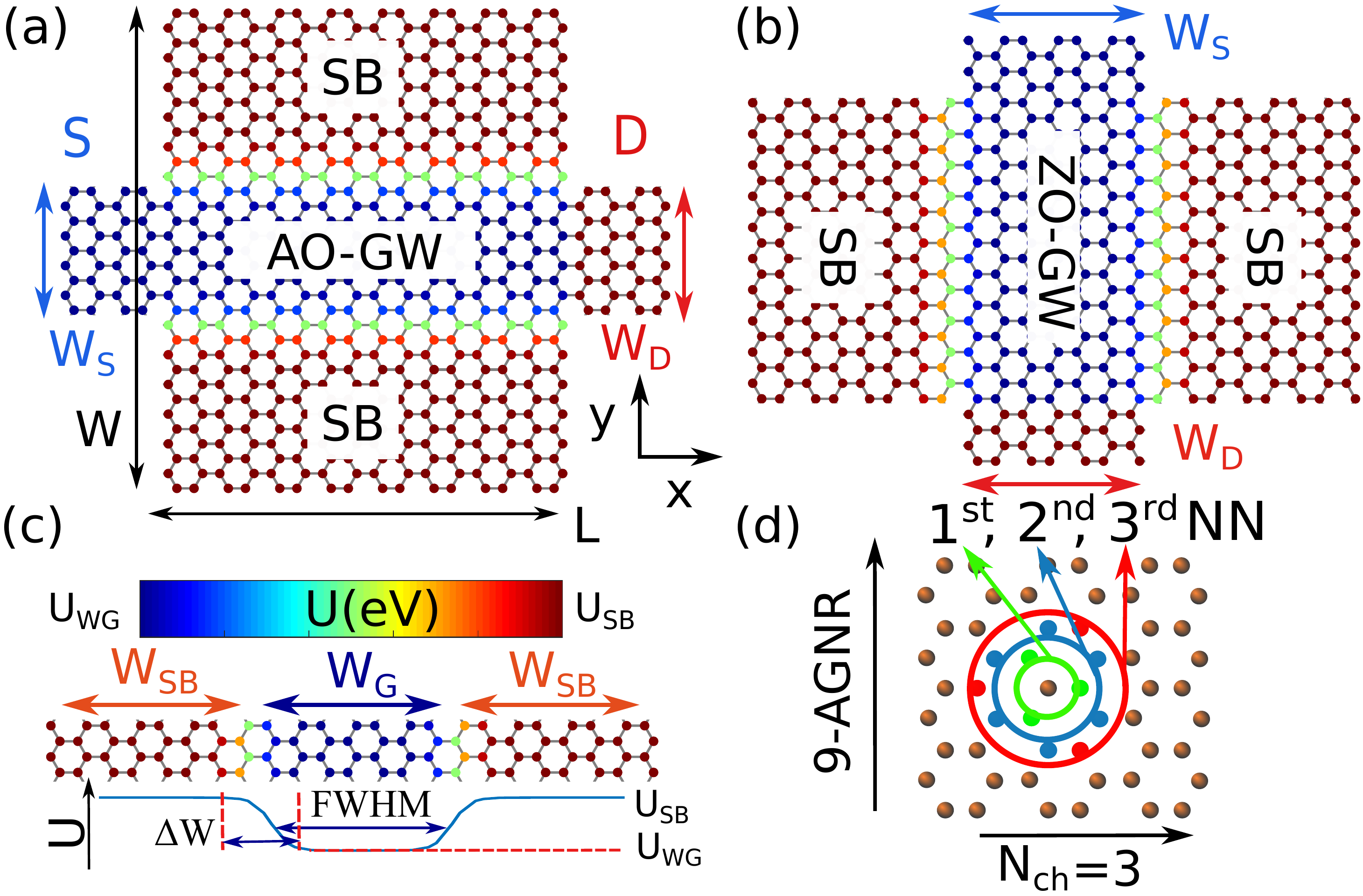}
\caption{(Color online) Schematic diagrams of graphene waveguides. (a) \proofreadtrue\changemarker{Armchair oriented waveguide (AO-GW). (b) Zigzag oriented waveguide (ZO-GW). SB indicates the side-barrier. (c) The cross section of ZO-GW showing the smooth variation of the on-site potential energy. The scale of on-site potential at each atomic site is indicated by different color. The potential profile (U) across the x-axis is shown underneath, which ranged from U$_{WG}$ on the bottom of the waveguide to U$_{SB}$ on the side-barriers. (d) An example of N$_A$-GNR with N$_A$~=~9 together with a small scattering area with N$_{ch}$~=~3 to show the different tight-binding approximations with 1st, 2nd, and 3rd nearest neighbors.}
}
	\label{fig1}
\end{figure}

Fig.~\ref{fig1} illustrates the geometry of our devices. Middle-size strips of graphene with width W and length L are considered as the scattering area, where the armchair and zigzag edges are distributed along the horizontal (x-axis) and vertical \proofreadtrue\changemarker{(y-axis)} directions, respectively. We introduce an external \proofreadtrue\changemarker{rectangular} gate to induce a spatially varied atomic on-site energy in \proofreadtrue\changemarker{the} graphene strip, which divides the scattering area into a \proofreadtrue\changemarker{centrally located} region of waveguide and two side-barriers. In this way, two distinct edge orientations for \proofreadtrue\changemarker{graphene waveguide} (AO-GW and ZO-GW) can be created as shown in Figs.~\ref{fig1}(a) and \ref{fig1}(b), respectively.
W$_G$ (W$_{SB}$) represents the width of waveguide (side-barrier) with fixed on-site energy \proofreadtrue\changemarker{U$_{WG}$ (U$_{SB}$),} in which we have \proofreadtrue\changemarker{considered} the full width at half maximum (FWHM) accounting for the smoothed on-site energy as shown in Fig.~\ref{fig1}(c). \proofreadtrue\changemarker{Note that the potential energy on the atomic sites is indicated by color in Figs. 1(a) and 1(b), and can be referred to the color bar shown in Fig. 1(c)}. Each \proofreadtrue\changemarker{graphene waveguide} contains two fundamental parts\proofreadtrue\changemarker{:} the  scattering area and leads (the areas that stick out from the scattering area). We use the notation N$_A$-AGNR to label the central scattering area. N$_A$ stands for the number of dimer lines and is defined as N$_A=1+\lfloor{W/(0.5\sqrt{3}a_{cc})}\rfloor$, in which W is the width of AGNR and a$_{cc}=0.142$~nm is the carbon-carbon bond length. The length of the scattering area (L) is related \proofreadtrue\changemarker{with} the chain number ($N_{ch}$) via N$_{ch}=\lfloor{L/(3a_{cc})}\rfloor$ (note that each chain contains 2N$_{A}$ atoms). \proofreadtrue\changemarker{Parameters N$_{A}$ and N$_{ch}$ are two essential inputs to build the scattering area.}

The second part of the device is contacts (source and drain) which are also made of carbon and are in fact finite-width GNRs attached to the scattering area, as illustrated by the extended GNRs sticking out of the rectangular \proofreadtrue\changemarker{region of} W$\times$L in Figs.~\ref{fig1}(a) and \ref{fig1}(b). 
\proofreadtrue\changemarker{The width of source (drain) in both orientations is labeled by W$_{S}$ (W$_{D}$) and is also related with the number of dimer lines in source (drain) by N$_{eS}$ (N$_{eD}$), where the first index (e =~a, z) stands for the edge orientation.} 
\proofreadtrue\changemarker{The orientation of scattering area is kept unchanged, whereas the position of the leads and the edge orientation of the waveguide are different for ZO-GW and AO-GW (see Figs.~\ref{fig1}(a) and \ref{fig1}(b))}. 
It can be assumed that wider leads (as compare to W$_G$) provide denser subbands and consequently higher density of state for carriers to get in and out of the waveguide. On the other hand, wider leads may also provide extra paths for carriers to go through \proofreadtrue\changemarker{the side-barriers} instead of the waveguide and thus the interference may demolish the coherent transmission from source to drain~\cite{mosallanejad2018coherent, mosallanejad2018perfectly}. Thus, in most configurations \proofreadtrue\changemarker{discussed in this report} W$_{S,~D}$ is equal to W$_{G}$ unless otherwise stated.
\begin{table}[]
	
	\centering
	\begin{tabular}{|l|l|l|l|l|l|l|l|}
		\hline
		approx. & $\epsilon_0(eV)$ & $t_0(eV)$ & $t_1(eV)$ & $t_2(eV)$ & $s_0(eV)$ & $s_1(eV)$ & $s_2(eV)$\\
		\hline
		1st & 0 & -2.74 & 0 & 0 & 0 & 0 & 0 \\
		\hline
		3rd & -0.36 & -2.78 & -0.12 & -0.068 & 0.106 & 0.001 & 0.003 \\
		\hline
	\end{tabular}
	
	\caption{\label{tab-tc}Hopping energies and overlap integral values for \proofreadtrue\changemarker{the} 1st (\proofreadtrue\changemarker{first row}) and the 3rd (second row) nearest neighbor \proofreadtrue\changemarker{tight-binding approximations~\cite{reich2002tight,kundu2011tight}}. 
	}
\end{table}

Moreover, our previous studies have shown that a metallic AGNR is a better choice to make an ideal contact to \proofreadtrue\changemarker{armchair oriented graphene waveguide}~\cite{mosallanejad2018perfectly,mosallanejad2018coherent}
. Indeed, the zero-energy modes in \proofreadtrue\changemarker{metallic AGNRs} \proofreadtrue\changemarker{permit} the low energy electrons from the source to be injected into \proofreadtrue\changemarker{the} waveguide region. The advantage of using metallic \proofreadtrue\changemarker{GNRs} as leads reflects itself as an early onset of the first conductance plateau around the Dirac point. Thus, we may modify N$_{aS,~aD}$ by 1 or 2 to yield \proofreadtrue\changemarker{a number of} dimer lines of \proofreadtrue\changemarker{N$_{aS,~aD}$~=~3$m$+2 ($m$ is an integer),} which is the condition for \proofreadtrue\changemarker{building} metallic \proofreadtrue\changemarker{AGNRs}. 
On the other hand, ideal ZGNR leads (with an even number of atoms in the unit cell) connecting to ZO-GWs do not need any modification, because they \proofreadtrue\changemarker{naturally} have zero\proofreadtrue\changemarker{-energy} modes.
Source leads have the same on-site energy as in the guiding region while the drain leads are grounded (zero on-site energy) in all examples. \proofreadtrue\changemarker{Tight-binding} Hamiltonian of a graphene device can be expressed as:
\begin{equation}
\label{eq:1}
H  =\sum_{i}{\mu_ic_i^\dagger c_i}+\sum_{i,j} {t_{i,j} (c_i^\dagger c_j)},
\end{equation}
where $c_i^\dagger (c_i)$ is the creation (annihilation) operator and $\mu_i$, indicates the on-site energy at the i-th atomic site. The on-site energy can be tuned through the external gate potentials and is \proofreadtrue\changemarker{described by} U as depicted in Fig.~\ref{fig1}(c). Hopping between \proofreadtrue\changemarker{the} nearest neighbors (e.g., i and j sites) is the origin of second term where $t_{i,j}$ denotes a fixed energy value based on \proofreadtrue\changemarker{ tight-binding approximations, as in Table~\ref{tab-tc}} ~\cite{reich2002tight,kundu2011tight}. A small size scattering area with N$_{ch}$~=~3 is shown in Fig.~\ref{fig1}(d) in which the 1st, 2nd\proofreadtrue\changemarker{,} and 3rd order \proofreadtrue\changemarker{tight-binding} approximations are indicated by \proofreadtrue\changemarker{green, blue} and red circles, respectively.
Following \proofreadtrue\changemarker{the} Landauer-B{\"u}ttiker formalism, conductance of a two-terminal device in low-temperature and low-bias can be expressed as $G=G_0T$, where $G_0~=~2e^2/h$ represents the quanta of conductance and \textit{T} is the transmission coefficient. Spin degree of freedom is included by the factor 2 in G$_0$ \proofreadtrue\changemarker{while} \textit{e} and \textit{h} are \proofreadtrue\changemarker{the electron charge and Planck's constant}~\cite{datta2005quantum}. \proofreadtrue\changemarker{The source-to-drain transmission coefficient \textit{T} can be calculated using the Caroli's formula}~\cite{caroli1971direct}: 
\begin{equation}
\label{eq:2}
T=trace(\Gamma_sG^r\Gamma_dG^a)\proofreadtrue\changemarker{,}
\end{equation}
where $\Gamma_{s}$ ($\Gamma_{d}$) is the broadening matrix of the source (drain) lead. $G^{r}$ ($G^a={G^r}^\dagger$) represents retarded \proofreadtrue\changemarker{(advanced) Green's} function given by
\begin{equation}
\label{eq:3}
G^r(E)=[(E+i\eta)S-H-\Sigma_{s}(E)-\Sigma_{d}(E)]^{-1}\proofreadtrue\changemarker{,}
\end{equation}
where $\eta$ is \proofreadtrue\changemarker{a} small infinitesimal number usually about $10^{-4}$. Here, $S$ is the \textit{overlap matrix} built in \proofreadtrue\changemarker{a similar way to the second term in Eq.~(\ref{eq:1}), and takes the form}
\begin{equation}
\label{eq:4}
S  =\sum_{i,j} {s_{i,j} (c_i^\dagger c_j)}\proofreadtrue\changemarker{,}
\end{equation}
where $s_{i,j}$ \proofreadtrue\changemarker{represents} the overlap integral between \proofreadtrue\changemarker{atomic orbitals (p$_z$)} located at \textit{i} and \textit{j}. It is worth noting that orbitals at two different atomic sites are not necessarily orthogonal to each other. \proofreadtrue\changemarker{Therefore, non-zero values exist on the \textit{S} matrix if the third (3$^{rd}$) nearest approximation is considered (see Table~\ref{tab-tc}).} \proofreadtrue\changemarker{However, these values are small due to the long-distance interactions between atomic orbitals.}
\proofreadtrue\changemarker{The} open boundary condition at the source and drain \proofreadtrue\changemarker{is} incorporated into the transport study via the last two terms in \proofreadtrue\changemarker{Eq.~(\ref{eq:3})}, which are the so called \textit{self-energy} terms. Self-energy matrices are calculated via $\Sigma_{s}=A_{s}^{\dagger}~g_{s}~A_{s}$ and $\Sigma_{d}=A_{d}~g_{d}~A_{d}^{\dagger}$, in which $A_{s,~d}$ are given by
\begin{equation}
\label{eq:5}
A_{s,~d}(E)=[(E+i\eta)S_{s\textbf{S},~\textbf{S}d}-H_{s\textbf{S},~\textbf{S}d}]\proofreadtrue\changemarker{.}
\end{equation}
Here, $H_{s\textbf{S}}$ and $S_{s\textbf{S}}$ are the \proofreadtrue\changemarker{\textit{interaction Hamiltonian}} and \textit{interaction overlap} matrices between the source and the first super cell in the scattering area, while $H_{\textbf{S}d}$ and $S_{\textbf{S}d}$ are the \textit{interaction Hamiltonian} and \textit{interaction overlap} matrices between the last supercell in the scattering area and drain lead (index \textit{\textbf{S}} refers to the scattering area whereas \textit{s} and \textit{d} denote the source and drain). In \proofreadtrue\changemarker{the} process of building $H_{s\textbf{S}}$ ($S_{s\textbf{S}}$), the \proofreadtrue\changemarker{\textit{i}-th index in} \proofreadtrue\changemarker{Eq.~(\ref{eq:1})} (\proofreadtrue\changemarker{Eq.~(\ref{eq:4})}) goes over \proofreadtrue\changemarker{the} atomic sites in the source lead while the \proofreadtrue\changemarker{\textit{j}-th} index goes over \proofreadtrue\changemarker{the atomic} sites in the first super-cells of \proofreadtrue\changemarker{the} central scattering area. $H_{\textbf{S}d}$ and $S_{\textbf{S}d}$ are constructed similarly. \proofreadtrue\changemarker{We employed the Sancho-Rubio iterative scheme to calculate the retarded \textit{surface Green's functions}}, $g_{s,~d}$~\cite{sancho1985highly, pourfath2014non}\proofreadtrue\changemarker{, from which one can} easily obtain \proofreadtrue\changemarker{the} \textit{broadening matrices} \proofreadtrue\changemarker{via} $\Gamma_{s,~d}=i( \Sigma_{s,~d}-\Sigma_{s,~d}^\dagger )$. Another important parameter is the local density of state (LDOS) given by
\begin{equation}
\label{eq:6}
LDOS(E)=(i/\pi)~diag(G^r(E)-G^a(E))\proofreadtrue\changemarker{,}
\end{equation}
where \textit{diag} refers to the diagonal elements of the matrix. We can also evaluate LDOS by extracting the real part of the diagonal elements of the spectral function $(G^r\Gamma_{s,~d}G^a)$. This parameter determines the spatial distribution of  wave function \proofreadtrue\changemarker{at a specific} Fermi energy. Inversion of the large matrix in \proofreadtrue\changemarker{Eq.~(\ref{eq:3}), which is} associated with the large number of atoms in \proofreadtrue\changemarker{the scattering area, is a massive task}. \proofreadtrue\changemarker{For many of the physical quantities such as the transmission function and LDOS, only part of the full Green’s function is required}. 
\proofreadtrue\changemarker{The recursive scheme, explained in detail in Ref.~\cite{thorgilsson2014recursive}, allows us to obtain the essential parts of the Green's function to perform the necessary calculations.}

In \proofreadtrue\changemarker{tight-binding} theory, expansion of free electron wave function in terms of the Block's wavefunction together with the minimization of energy converts the Schr\"odinger equation into an eigenvalue matrix equation, H($\mathbf{k}$)-E($\mathbf{k}$)S($\mathbf{k}$)=0, where $\mathbf{k}$ is the two dimensional wavevector whose range is determined by high symmetry points in graphene's reciprocal lattice~\cite{saito1998physical}. In systems with a physical confinement in the transverse direction, it is possible to further simplify the 2D bandstructure calculation by assuming a plane-wave wavefunction in the longitudinal direction\proofreadtrue\changemarker{:} {\fontsize{14}{13}\selectfont $e^{ik_\parallel x_\parallel}$}\proofreadtrue\changemarker{,} where the index {\fontsize{11}{12}\selectfont $\parallel$} denotes the \proofreadtrue\changemarker{longitudinal (transport)} direction. Physical confinement in \proofreadtrue\changemarker{the} transverse direction leads to H(${k_\perp}$)-E(${k_\perp}$)S(${k_\perp}$)=0\proofreadtrue\changemarker{, where the index {\fontsize{11}{12}\selectfont $\perp$} denotes the transverse direction}. \proofreadtrue\changemarker{The eigenvalues E(${k_\perp}$) of the following characteristic equation (the so-called \textit{secular equation}),
	\begin{equation}
	\label{eq:7}
	det(H({k_\perp})-E({k_\perp})S({k_\perp}))=0, 
	\end{equation}
	give rise to the quasi-one dimensional band structure}. Note that $H({k_\perp})$ is \proofreadtrue\changemarker{given} by 
\begin{equation}
\label{eq:8}
H({k_\perp})\equiv H_{lc}e^{(-ik_\perp a_c)}+H_{cc}+H_{cr}e^{(ik_\perp a_c)}\proofreadtrue\changemarker{,}
\end{equation}
where $a_c$ is the distance between the neighbor super-cells.
\proofreadtrue\changemarker{$H_{cc}$ denotes the interaction Hamiltonian between all atoms in the central super-cell, while  $H_{lc~(cr)}$ represents the interaction Hamiltonian between atoms in the left (central) super-cell with atoms in the central (right) super-cell}. 
One can use \proofreadtrue\changemarker{Eq.~(\ref{eq:1})} to build each of \proofreadtrue\changemarker{the} Hamiltonian matrices \proofreadtrue\changemarker{in Eq.~(\ref{eq:8})}.
\proofreadtrue\changemarker{$S({k_\perp})$ has a similar form to $H({k_\perp})$ in which $S_{lc}$, $S_{cc}$ and $S_{cr}$ (constructed via Eq.~(\ref{eq:4})) replacing the equivalent Hamiltonian terms in Eq.~(\ref{eq:8})}. 
\proofreadtrue\changemarker{Altogether, Eq.~(\ref{eq:7}) can be constructed to solve the eigenvalue problem.}

\section{III. Results and Discussions}

\subsection{A. Straight Waveguides}
\proofreadtrue\changemarker{We begin our study by considering straight graphene waveguides in both edge orientations (AO-GWs and ZO-GWs), exploring three different side-barrier widths (W$_{SB}$) and investigating the effect of W$_{SB}$ on the conductance.}
The length of \proofreadtrue\changemarker{graphene waveguide} (L) and the width of the guiding region (W$_{G}$) are fixed at 100~nm and 20~nm, respectively. 
\proofreadtrue\changemarker{The 20~nm wide guiding region is equivalent to the number of dimer lines  N$_{A}$-GW~=~163 in AO-GW and N$_{Z}$-GW~=~188 in ZO-GW.}
\proofreadtrue\changemarker{The total width of scattering area W is 40, 60 and 80~nm which corresponds to W$_{SB}$~=~10, 20 and 30~nm, respectively.} 
At the same time\proofreadtrue\changemarker{, leads with the number of dimer lines N$_{{aS},~{aD}}$~=~161 (metallic armchair leads; \textit{a} stands for armchair and \textit{S}(\textit{D}) stands for source (drain)) and N$_{{zS},~ {zD}}$~=~188 (symmetric zigzag leads; \textit{z} stands for zigzag) have been considered for AO-GW and ZO-GW, respectively.
} 
\proofreadtrue\changemarker{The on-site potential energy in the} scattering area is smoothly varied within $\Delta$W~=~44a$_{cc}$~$\approx$~6.25~nm from \proofreadtrue\changemarker{U$_{SB}$~= 0~eV} at the side-barriers to \proofreadtrue\changemarker{U$_{WG}$~=~-0.3~eV} at the guiding area for all devices~\cite{mosallanejad2018coherent}.
\proofreadtrue\changemarker{As mentioned earlier, source leads and waveguide areas set to possess the same potential energy (U$_{WG}$) while drain leads are grounded in all samples}.
\begin{figure}
	\centering
	\includegraphics[width=8.6cm]{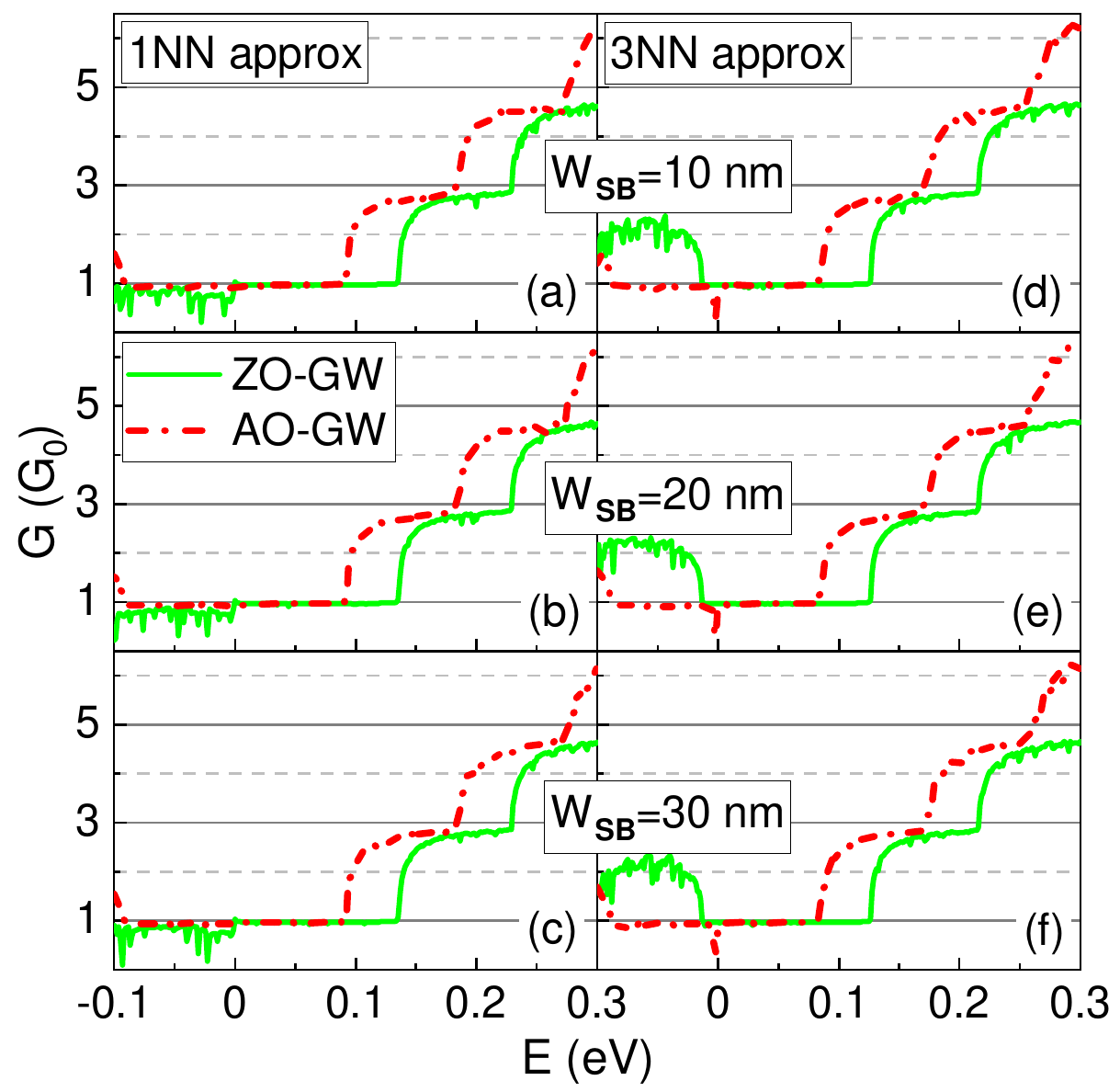}
	
	\caption{Conductance of 20~nm-wide AO-GW (red-dot lines) and ZO-GW (green-solid lines) \proofreadtrue\changemarker{with different} side-barrier width, W$_{SB}$. (a)-(c) with \proofreadtrue\changemarker{the} 1NN approximation and (d)-(f) with the 3NN approximation.}
	\label{fig2}
\end{figure}
\proofreadtrue\changemarker{We conducted a transport study for these six samples by considering  both the first (1NN) and the third (3NN) nearest tight-binding approximations. The results are} shown separately on the left and right panels in Fig.~\ref{fig2}. 
\proofreadtrue\changemarker{In Fig.~\ref{fig2} both the 20~nm ZO-GW and AO-GW exhibit a quantization of conductance G~=~1, 3, 5 G$_0$ in each configuration (see the green curve and red curve in each panel).} 
The first plateau of ZO-GW is clearly wider in energy axis than that of AO-GW. 
\proofreadtrue\changemarker{The first conductance plateaus for both ZO-GW and AO-GW are flat, whereas other higher plateaus are not, and show a gradual losing flatness toward more positive energies.} 
\proofreadtrue\changemarker{Importantly, the effect of side-barrier widths (W$_{SB}$)} seems negligible for both orientations.
\proofreadtrue\changemarker{This suggests a minimum influence of edge disorders on conductance of a gate-defined graphene waveguide as long as the edges (the border between side-barriers and vacuum) are far enough from the waveguide area.} 
\proofreadtrue\changemarker{When the 3NN approximation is
employed, noticeable dips in the conductance of AO-GWs (red-dot lines) appeared around E~=~0~eV, as can be seen in Figs.~\ref{fig2}(d)-(f).} 
This can be understood by the fact that the 3NN approximation tends to yield a small band gap in an AGNR (i.e., terminals)~\cite{wang2016energy}. 
Both 1NN and 3NN \proofreadtrue\changemarker{approximations} give rise to the noisy conductance features at E$<$0~eV in ZO-GW. 
\proofreadtrue\changemarker{Similar noisy conductance has also been observed in a AO-GW but at much lower energy levels. For example, E$<$-0.2~eV (not shown in Figs.~\ref{fig2})~\cite{mosallanejad2018perfectly}.}
We attribute these noises to the increase of current passing through side-barriers in this range of energy. \proofreadtrue\changemarker{At higher energies, the plateaus gradually disappear because there are only a few confined wavefunctions localized in the waveguide area.}

\begin{figure}
	\centering
	\includegraphics[width=8.6cm]{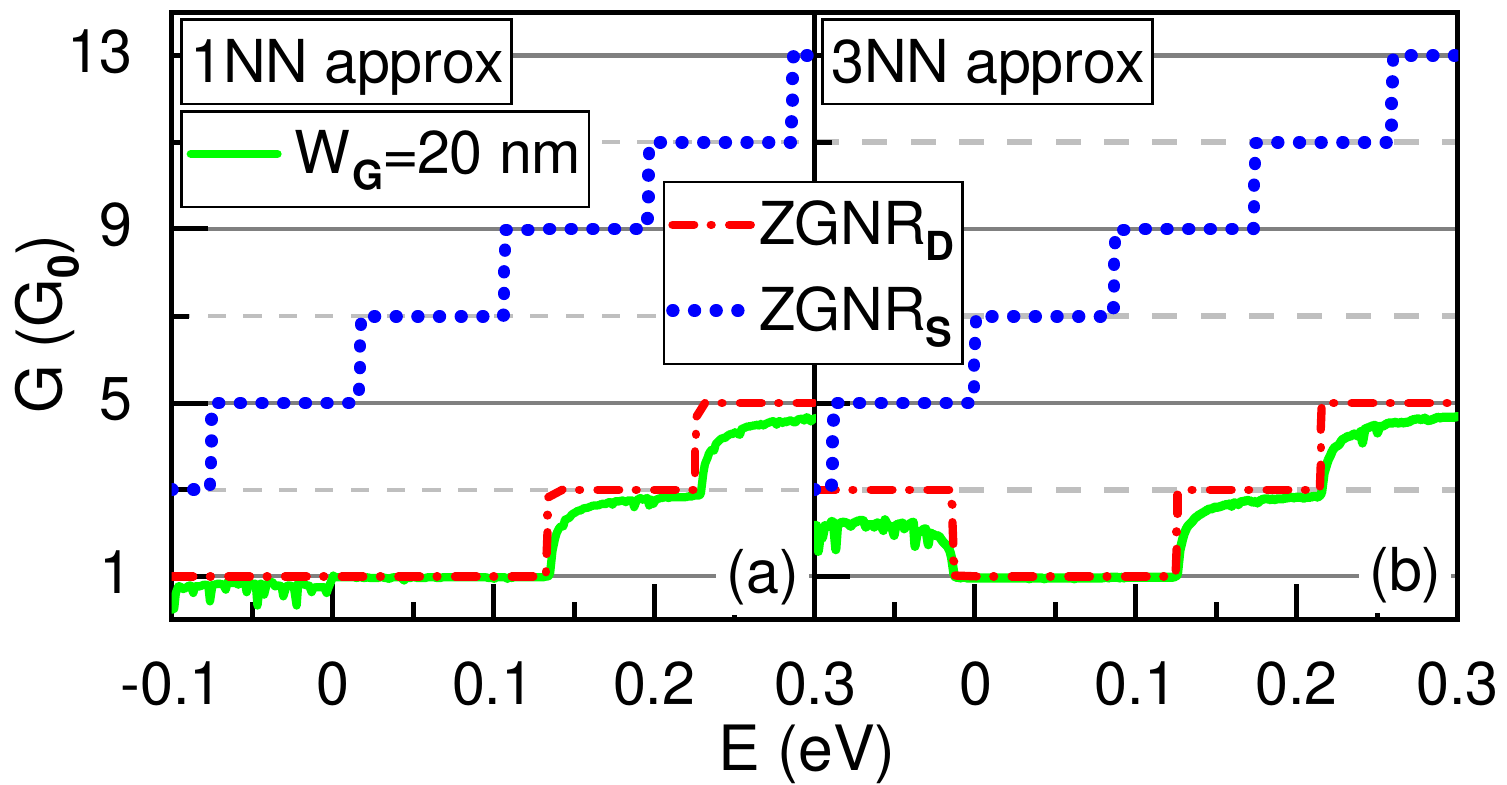}
	
	\caption{Conductance of source (blue-dot line), drain (red-dashed-dot line) and ZO-GW (green-solid line) considering (a) 1NN and (b) 3NN \proofreadtrue\changemarker{approximations}.}
	\label{fig3}
\end{figure}

In addition, with the 3NN approximation, the difference in conductance between two orientations became more visible. \proofreadtrue\changemarker{For example,} the conductance of ZO-GW exhibits larger values at E$<$0~eV. 
This difference can be explained by comparing the conductance of the drain electrode and \proofreadtrue\changemarker{ZO-GW} under the 3NN approximation, as shown in Fig.~\ref{fig3}. 
Note that, the conductance of \proofreadtrue\changemarker{the drain electrode refers to the conductance of the semi-infinite GNR that is used as a drain lead in our structure.}  
The correspondence between the red-dashed-dot line and green-solid line in Fig.~\ref{fig3} suggests that the conductance of \proofreadtrue\changemarker{the waveguide} follows the conductance behavior of the drain terminal.
 
\begin{figure}[]
	\centering
	\includegraphics[width=8.6cm]{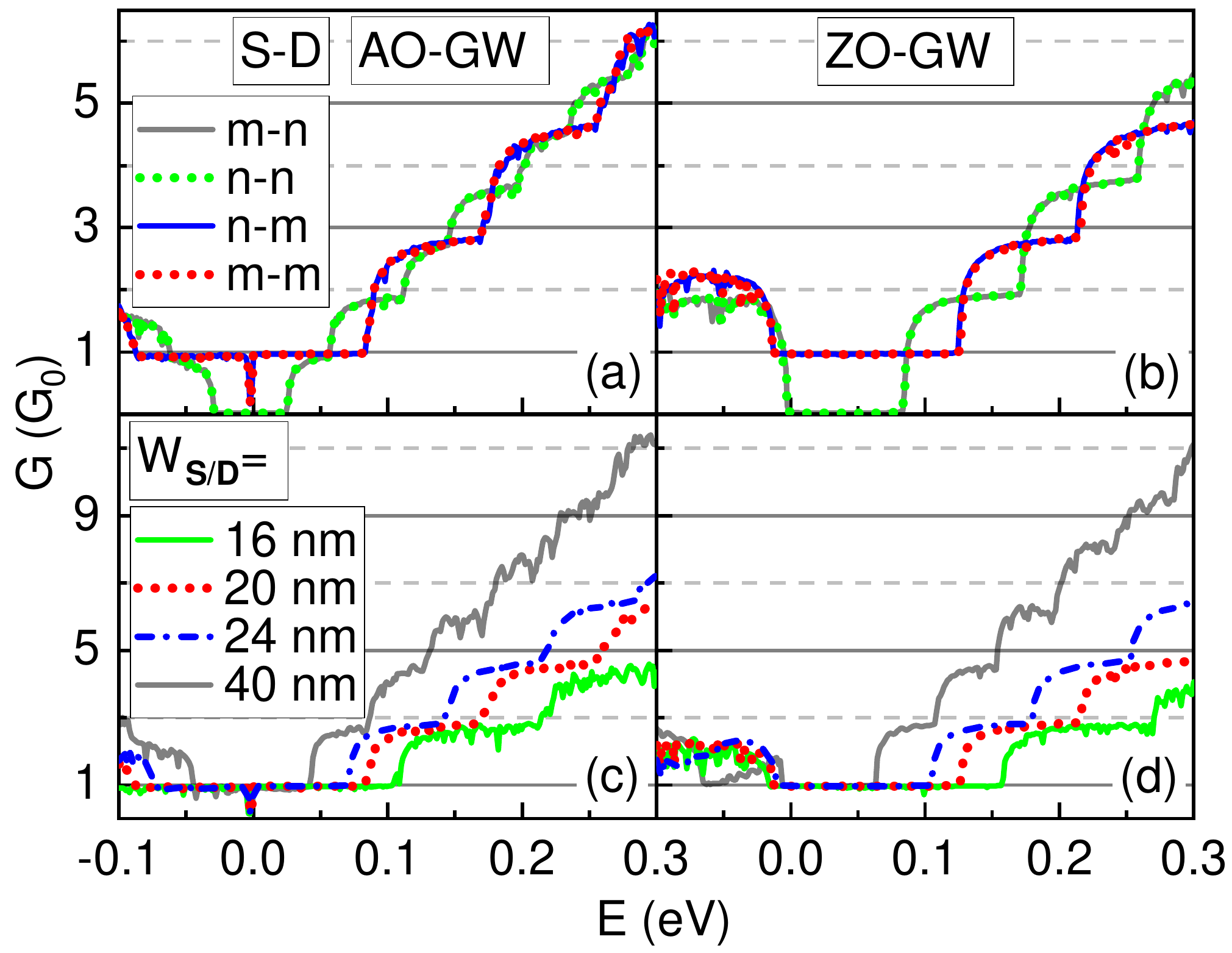} 
	\caption{ \proofreadtrue\changemarker{The conductance through straight graphene waveguides considering W$_{SB}$=~20~nm and W$_G$=~20~nm with different combination of leads nature. (a) Conductance of AO-GW for different combination of metallic (m) and nonmetallic (n) leads. (b) Same as (a) for ZO-GW. (c) Conductance of AO-GW for various widths of metallic leads. (d) Same as (c) for ZO-GW.}
	}
	\label{fig4}
\end{figure}

We further explore the effect of \proofreadtrue\changemarker{leads on waveguide transport properties}. Here, we modified \proofreadtrue\changemarker{the number of dimer lines of} leads by 1 or 2 to make \proofreadtrue\changemarker{them} either metallic or semiconducting (nonmetallic). In contrast to the insensitivity of conductance to the widths of side-barriers, \proofreadtrue\changemarker{conductance of waveguide for both orientations shows a clear dependence on the metallic (m) or nonmetallic (n) nature of leads, as shown in \proofreadtrue\changemarker{Figs}.~\ref{fig4}(a) and \ref{fig4}(b).
}
One third of \proofreadtrue\changemarker{the} AGNRs and an ideal ZGNR have metallic behavior because their band structures possess \proofreadtrue\changemarker{zero-energy} mode.

Different combinations of metallic and non-metallic leads are considered for a previously studied configuration, \proofreadtrue\changemarker{i.e.,} W$_{SB}$~=~20~nm and W$_{G}$~=~20~nm. 
Non-metallic drain in AO-GW yields a finite gap on conductance around E~=~0~eV (gray-solid and green-dot lines in Fig.~\ref{fig4}(a)) while the conductance of a configuration with \proofreadtrue\changemarker{ non-metallic source and metallic drain is identical to that with both metallic leads (i.e., blue-solid line is identical to red-dot line in Fig.~\ref{fig4}(a)).}
Moreover, the conductance of AO-GW with non-metallic drain (m-n and n-n) \proofreadtrue\changemarker{shows shorter spacing between plateaus with quantization steps at multiple of G$_0$ compared to that with metallic drain (n-m and m-m) which shows quantization steps at multiple of 2G$_0$.}
\proofreadtrue\changemarker{On the other hand, an ideal ZGNR (with closed hexagonal crystal structure) represented by an even number of dimer lines is indeed metallic. However, a ZGNR with an odd number of dimer lines results in breaking the crystal symmetry and is non-metallic due to the absence of the zero-energy mode.}
 \proofreadtrue\changemarker{As a result, the gap in conductance is even wider in the case of ZO-GW with disordered (non-metallic) drain (green-dot line in Fig.~\ref{fig4} (b)).} 
 Here, we refer a ZGNR lead with \proofreadtrue\changemarker{an} odd number \proofreadtrue\changemarker{of} dimer lines as a disordered lead. Also, like AO-GW, the conductance of ZO-GW with \proofreadtrue\changemarker{a} non-metallic source and  \proofreadtrue\changemarker{a} metallic drain (n-m) is identical to that with both metallic leads (m-m), as shown by the blue-solid and red-dot lines in Fig.~\ref{fig4} (b). \proofreadtrue\changemarker{Configurations with non-metallic drain (m-n and n-n) in ZO-GW do} not change the quantization step (in contrast to AO-GW) but it has shifted the conductance both vertically and horizontally, as shown in Fig.~\ref{fig4} (b). This result again indicates that the nature of the drain plays a significant role on the conductance \proofreadtrue\changemarker{of graphene waveguide for} both orientations. 
 Therefore, we adopted metallic leads for the rest of our studies because they yield early onset of \proofreadtrue\changemarker{non-zero conductance plateau for both edge orientations}.
 \proofreadtrue\changemarker{Altering the width of leads at nanometer scale also influences the conduction of graphene waveguide, as shown in Figs.~\ref{fig4}(c) and \ref{fig4}(d) for both edge orientations.}
 Wider conductance plateaus are presented for short leads and vice versa. 
 \proofreadtrue\changemarker{Note that the} situation W$_{D,~S}\neq$~W$_G$ has added a visible level of noise to the conductance plateaus \proofreadtrue\changemarker{in the cases of} much shorter (16~nm) and much wider (40~nm) leads as compared to the primary case of W$_{D,~S}=$~W$_G$.

\begin{figure*}
	\centering 
	\includegraphics[width=17cm]{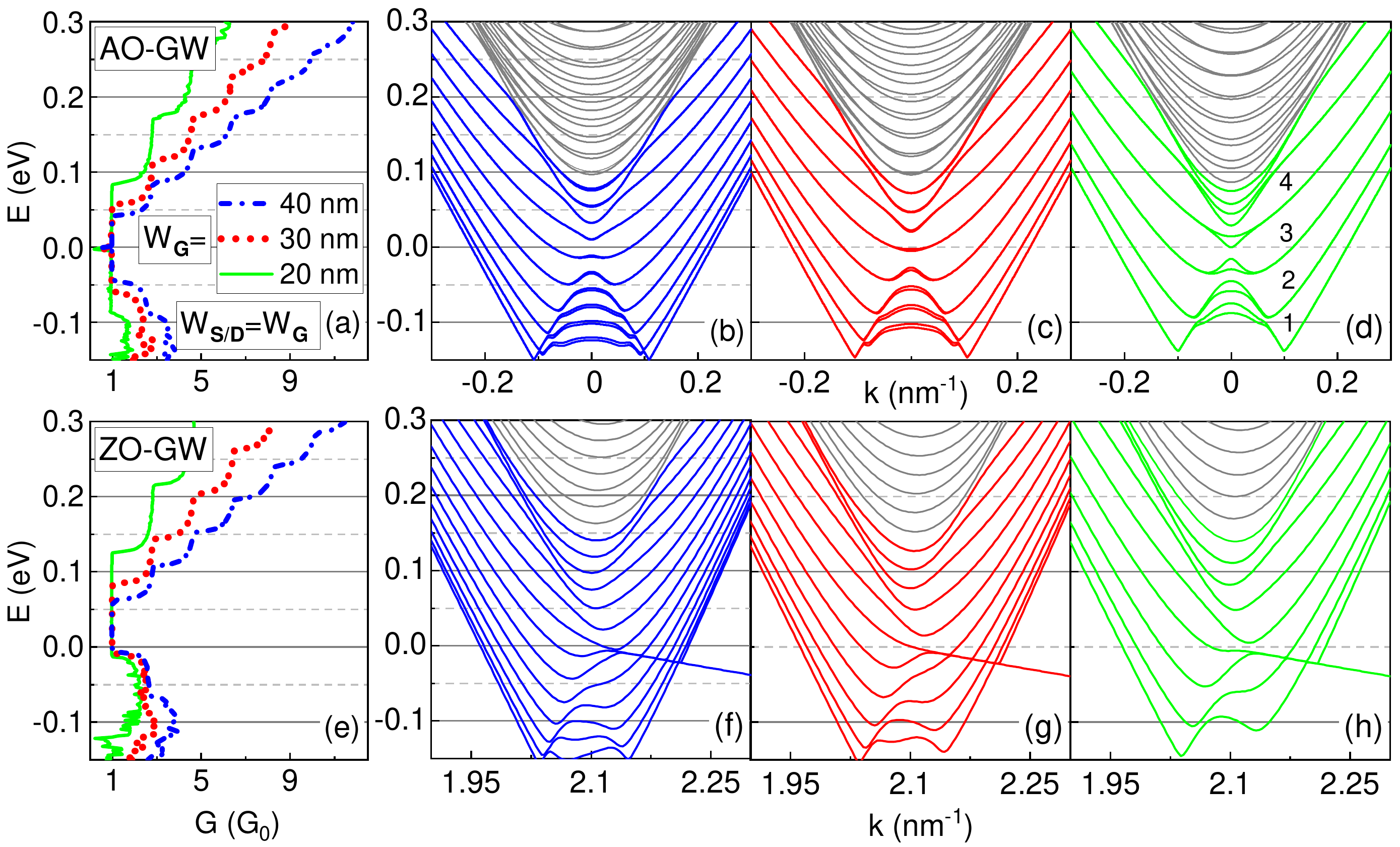}
	\caption{(Color online) \proofreadtrue\changemarker{(a) Conductance of AO-GWs for various W$_G$~=~20, 30 and 40~nm. (b)-(d) Band structure plotted for different W$_G$  employed in (a). The color of the bands in each panel correspond to the color used in (a). (e) Conductance of ZO-GWs for various W$_G$ used in (a). (f)-(h) band structure plotted for corresponding W$_G$ of (e). Solid gray lines in the band structures denote the bands corresponding to the wavefunctions that are not confined in the waveguide.} }
	\label{fig5}
\end{figure*}

\proofreadtrue\changemarker{In further study of the effect of parameter W$_G$ on the conductance of graphene waveguide for both edge orientations, we evaluated three values of W$_G$~(20, 30 and 40~nm), with leads satisfying the condition W$_{D,~S}$~=~W$_G$. For these tests, length L~=~100~nm and side-barriers W$_{SB}$~=~20~nm are kept fixed.} 
\proofreadtrue\changemarker{Conductance} and quasi-one dimensional band structures \proofreadtrue\changemarker{for} supercells corresponding to each W$_G$ are plotted in Fig.~\ref{fig5} with the same color schemes. 
For both edge orientations, as W$_G$ decreases from 40~nm to 20~nm, conductance plateaus get longer. \proofreadtrue\changemarker{This is a} result of larger spacing between the energy bands, as \proofreadtrue\changemarker{visible in Figs.~\ref{fig5}(b)-\ref{fig5}(d) and \ref{fig5}(f)-\ref{fig5}(h)}.
Note that \proofreadtrue\changemarker{the} subbands of AO-GW are two-fold degenerate \proofreadtrue\changemarker{(see Fig.~\ref{fig5}(b)-\ref{fig5}(d))} while \proofreadtrue\changemarker{the} subbands of ZO-GW are not degenerate \proofreadtrue\changemarker{(see Figs.~\ref{fig5}(f)-\ref{fig5}(h))}. 
\proofreadtrue\changemarker{For a specific W$_G$,} one can deduce that the first plateau on conductance for E$>$0~eV \proofreadtrue\changemarker{does not originate} from the first band of \proofreadtrue\changemarker{the graphene waveguide} by tracking the background gray dashed lines between the conductance and \proofreadtrue\changemarker{the corresponding bands} in Fig.~\ref{fig5}. 
For instance, {the fourth band of the} 20~nm wide \proofreadtrue\changemarker{graphene waveguide} in Fig.~\ref{fig5}(d) (bands are named by numbers regardless of degeneracy) around E~=~0~eV coincides with the beginning of the first plateau in Fig.~\ref{fig5}(a) (solid-green line).
To further explore the transport properties of \proofreadtrue\changemarker{graphene waveguide} in two different edge orientations, local density of states (LDOS) are calculated for the case of W$_G$~=~20~nm.

\begin{figure}
	\includegraphics[width=8.8cm]{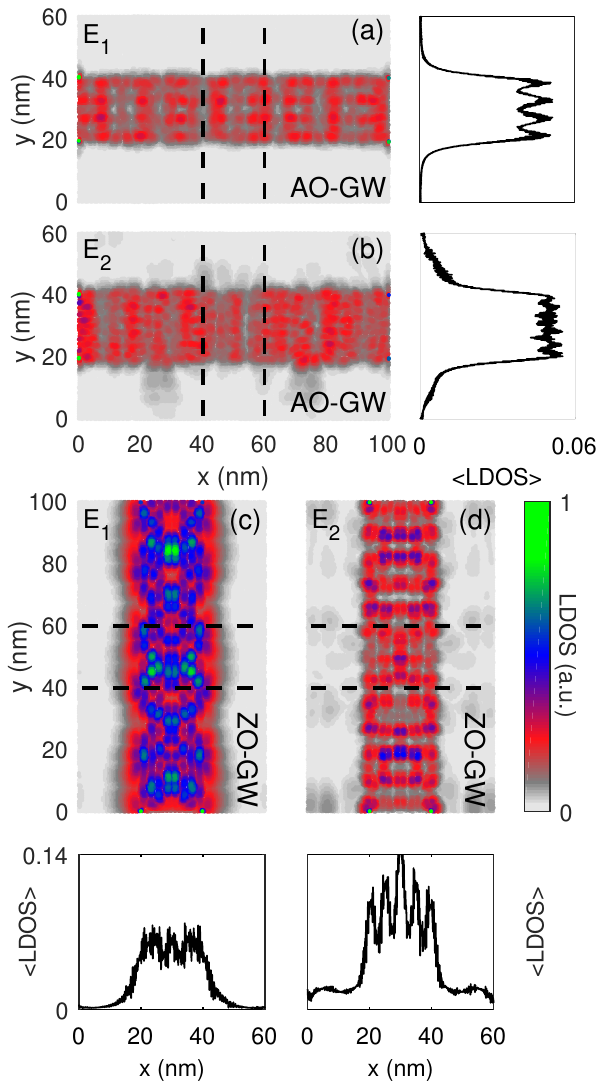}
	\caption{(Color online) \proofreadtrue\changemarker{LDOS for a AO-GW (a) at E$_1$~=~0.05 eV and (b) at E$_2$~=~0.15 eV. Right panels of (a) and (b) show the averages of the unnormalized LDOSs ($<$LDOS$>$) in the selected region between the black-dashed lines shown in (a) and (b). (c) and (d) The same as (a) and (b) but for ZO-GW.}   
		}
	\label{fig6}
\end{figure}

In Figs.~\ref{fig6}(a)-\ref{fig6}(d), normalized LDOS for both \proofreadtrue\changemarker{orientations of the 20~nm wide waveguide } are presented at two Fermi energies \proofreadtrue\changemarker{(E$_1$~=~0.05~eV and E$_2$~=~0.15~eV)}, which correspond to \proofreadtrue\changemarker{the conductance plateau at}  G$_0$ and 3G$_0$, respectively. 
\proofreadtrue\changemarker{Right (lower) panels of Figs.~\ref{fig6}(a)-\ref{fig6}(b) (Figs.~\ref{fig6}(c)-\ref{fig6}(d)) plot the average of the unnormalized LDOS ($<$LDOS$>$) within the black-dashed lines shown in Figs.~\ref{fig6}(a)-\ref{fig6}(b) (Figs.~\ref{fig6}(c)-\ref{fig6}(d)).}
Reasonable localization of LDOS is \proofreadtrue\changemarker{apparent} within the waveguide area at E$_1$ for both \proofreadtrue\changemarker{orientations}, as shown in Figs.~\ref{fig6}(a) and \ref{fig6}(c). 
The four peaks visible \proofreadtrue\changemarker{in the right panel of Fig.~\ref{fig6}(a) correspond} to the fourth mode in the band structure of AO-GW (see Fig.~\ref{fig5}(d)), which contributes to the first plateau \proofreadtrue\changemarker{in the W$_G$~=~20~nm waveguide}. 
Similar analysis can be performed for other \proofreadtrue\changemarker{graphene waveguides} with different widths and at different energies. Comparison between Fig.~\ref{fig6}(a) and Fig.~\ref{fig6}(b) (or Figs.~\ref{fig6}(c) and Fig.~\ref{fig6}(d)) shows stronger confinement of wavefunction at E$_1$ as compared to E$_2$.
Nevertheless, \proofreadtrue\changemarker{$<$LDOS$>$ shows that leakage of wavefunction toward side-barriers is still negligible at E$_2$ for both edge orientations}. 

\subsection{B. U-, L-Shape and Split Waveguides}
\begin{figure}
	\includegraphics[width=8.6cm]{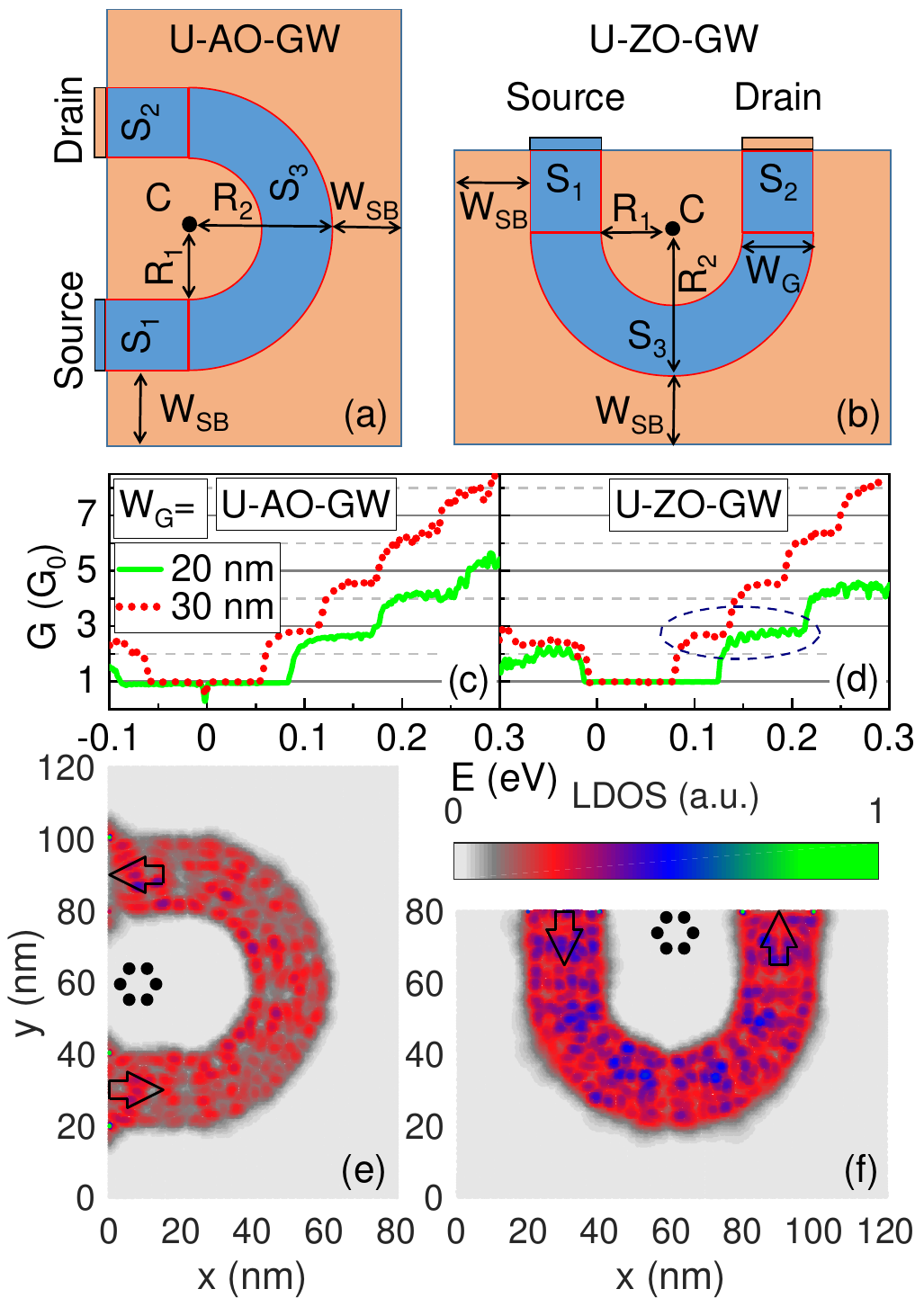} 
	\caption{(Color online) \proofreadtrue\changemarker{(a) and (b) show the schematic diagram for U-AO-GW and U-ZO-GW. (c) and (d) show the conductance of U-AO-GW and U-ZO-GW, with W$_G$~=~20~nm (red-dot line) and 30~nm (green-solid line), respectively. (e) and (f) show LDOS calculated for U-AO-GW and U-ZO-GW with W$_G$~=~20~nm and at E~=~0.03~eV.}}
	\label{fig7}
\end{figure}

\proofreadtrue\changemarker{In this section, we further study the transport properties of waveguides with the geometries that can be potentially used in nanoelectronic devices.}
\proofreadtrue\changemarker{Three types of curved waveguides, U-shape, L-shape and split-shape, have been taken into account to investigate the conductance profile and the ability to confine the charge carriers in these highly bent structures.} 
\proofreadtrue\changemarker{In a U-shape graphene waveguide, both the source and drain leads are connected to the same edge orientation (either armchair or zigzag edge).} 
\proofreadtrue\changemarker{In the following, we use the notation U-AO-GW (U-ZO-GW) to represent the U-shape AO-GW (ZO-GW). A U-AO-GW (U-ZO-GW) can be constructed by bending a straight AO-GW (ZO-GW) by 180$^\circ$ as shown in Fig.~\ref{fig7}(a) (Fig.~\ref{fig7}(b)).}
\proofreadtrue\changemarker{Dimension of the scattering area is W$\times$L~=~120~nm$\times$80~nm for U-AO-GW and W$\times$L~=~80~nm$\times$120~nm for U-ZO-GW. Here, we consider the waveguides with two different widths ( W$_G$~=~20~nm and 30~nm) in each orientation.}
The width of \proofreadtrue\changemarker{the} middle-barrier between \proofreadtrue\changemarker{the} source and drain \proofreadtrue\changemarker{(i.e., 2R$_1$ in Fig.~\ref{fig7}(a) and Fig.~\ref{fig7}(b))} is set to 40~nm (30~nm) when W$_G$~=~20~nm (30~nm), while W$_{SB}$~=~20~nm \proofreadtrue\changemarker{was consistent across all structures}. 
\proofreadtrue\changemarker{The on-site potential energy of the U-shape waveguides with W$_G$~=~20~nm is constructed by a combination of three segments: two AO(ZO)-GWs with L~=~20~nm which are parallel to each other, and half of a circular waveguide with inner (outer) radius of 20~nm (40~nm) which provides a smooth bending around the center of the circular part (i.e., point C in Fig.~\ref{fig7}(a) and Fig.~\ref{fig7}(b))}.
\proofreadtrue\changemarker{Conductance of the U-AO-GWs and the U-ZO-GWs both resemble that of their counterparts (straight AO-GWs and ZO-GWs), as can be observed by comparing Fig.~\ref{fig7} (c) with Fig.~\ref{fig5}(a) and Fig.~\ref{fig7}(d) with Fig.~\ref{fig5}(e).}
In the U-shape case, the general form of quantized conductance is preserved, but the second plateau is modulated by a visible oscillation as highlighted by a dashed ellipse in Fig.~\ref{fig7}(d). This oscillation is more pronounced \proofreadtrue\changemarker{in the W$_G$~=~20~nm case and becomes less visible when W$_G$ is 30~nm.} 
\proofreadtrue\changemarker{The normalized LDOS for U-shape waveguide with W$_G$~=~20~nm in both orientations at a given energy of E~=~0.03~eV (which
locates within the first plateau) is plotted in Figs.~\ref{fig7}(e) and \ref{fig7}(f), respectively.}
\proofreadtrue\changemarker{Both LDOS again show reasonable confinement at given Fermi energy which corresponds to the conductance plateau.}

\begin{figure}[]
	\includegraphics[width=8.75cm,height=11cm]{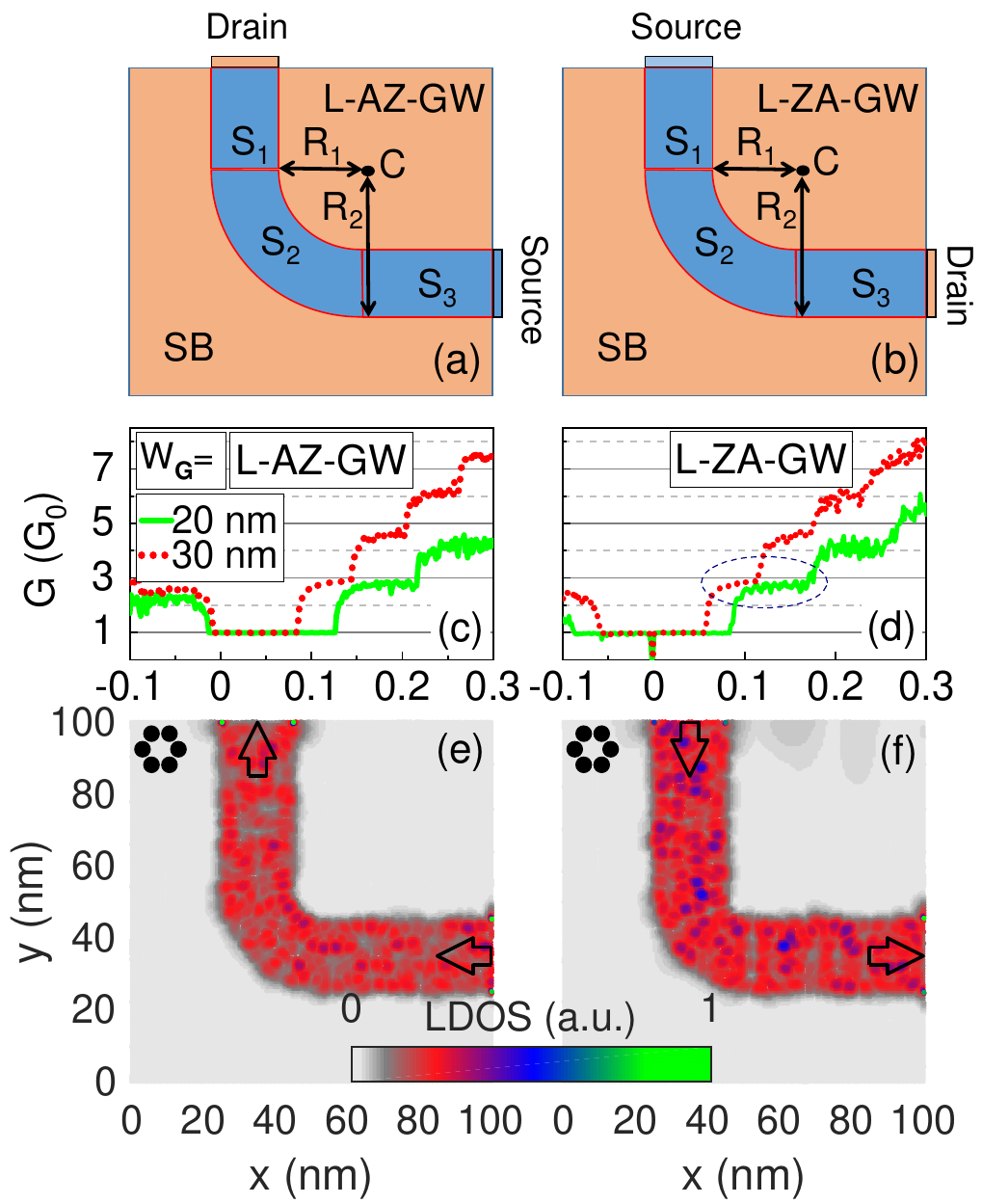}
	\caption{(Color online) Schematic diagram \proofreadtrue\changemarker{for} (a) L-AZ-GW and (b) L-ZA-GW. (c) and (d) show the \proofreadtrue\changemarker{conductance} of L-AZ-GW and L-ZA-GW, with W$_G$~=~20~nm (red-dot line) and 30~nm (green-solid line), respectively. \proofreadtrue\changemarker{(e) and (f) show LDOS calculated for L-AZ-GW and L-ZA-GW with W$_G$~=~20~nm and at E~=~0.05~eV.}}
	\label{fig8}
\end{figure}

Next, \proofreadtrue\changemarker{we studied the L-shape graphene waveguide to investigate the effect of $90^\circ$ bending on their transport properties}.
Here, we considered two configurations of L-shape waveguide in a fixed-size scattering area (W~=~L~=~100~nm), as shown in Fig.~\ref{fig8}(a) and Fig.~\ref{fig8}(b). 
\proofreadtrue\changemarker{First, a AO-GW bent to become a ZO-GW, with source on the zigzag interface and drain on the armchair interface, as labeled as L-AZ-GW. Secondly, a ZO-GW bent to become a AO-GW, with source on the armchair interface and drain on the zigzag interface, as labeled as L-ZA-GW.}
\proofreadtrue\changemarker{Note that the edge orientation of the scattering area is fixed while the location of source and drain leads is different for each case, as visible in Fig.~\ref{fig8}(a) and Fig.~\ref{fig8}(b).}
\proofreadtrue\changemarker{The waveguide (equivalently the on-site potential energy) is constructed using a combination of AO-GW and ZO-GW (both with L~=~50~nm) perpendicular to each other, and a quarter of a circular waveguide with inner (outer) radius of 10~nm (30~nm), which provides a smooth $90^\circ$ bending around the center of the system (i.e., point C in Figs.~\ref{fig8}(a) and \ref{fig8}(b)).}
\proofreadtrue\changemarker{To calculate the conductance of the aforementioned configurations, one only needs to switch the on-site potential energy between source and drain, and the relative positions of $\Gamma_{s}$ and $\Gamma_{d}$ in Eq.~(\ref{eq:2}).}
\proofreadtrue\changemarker{Conductance of the L-shape waveguide in each configuration, with W$_G$~=~20~nm and 30~nm, is plotted in Figs.~\ref{fig8}(c) and \ref{fig8}(d), respectively.}
\proofreadtrue\changemarker{Consistent with the previous results of straight waveguides, conductance of the L-shape graphene waveguides (both ZA and AZ) show dependence on the nature of the drain, as can be seen by comparing  Fig.~\ref{fig8}(c) with Fig.~\ref{fig5}(a) and Fig.~\ref{fig8}(d) with Fig.~\ref{fig5}(e) for W$_G$~=~20~nm and 30~nm}. 
Conductance of a 20~nm L-ZA-GW also shows a visible oscillation at the second conductance plateau, \proofreadtrue\changemarker{which is similar to the case of U-shape graphene waveguide}. This phenomenon could be attributed to the \proofreadtrue\changemarker{bending-induced} scattering between \proofreadtrue\changemarker{$K$ and $K^{\prime}$ sub-lattices}. 
Similarly, we \proofreadtrue\changemarker{calculated the LDOS of L-shape graphene waveguides} with W$_G$~=~20~nm and at E~=~0.05~eV (\proofreadtrue\changemarker{within} the first conductance plateau). Both \proofreadtrue\changemarker{L-shape graphene waveguides present a decent} confinement of wave function along the straight parts and around the bending area, \proofreadtrue\changemarker{as shown} in Figs.~\ref{fig8}(e) and \ref{fig8}(f). 
As an extension to \proofreadtrue\changemarker{the L-shape graphene waveguide, we subsequently studied} the split waveguides, which could be viewed as the counterpart of an optical beam splitter. 
\proofreadtrue\changemarker{The on-site energy of a split graphene waveguide can be constructed by combining that of two adjacent L-shape waveguides bent in opposite directions.} 
\proofreadtrue\changemarker{The split waveguide built in the scattering area consists of two parts: a stem part and two split parts}. 
 In our example, \proofreadtrue\changemarker{the stem part is 40~nm wide} and it splits equally into two 20~nm wide \proofreadtrue\changemarker{bent graphene waveguides}. 
\begin{figure}
\includegraphics[width=8.6cm]{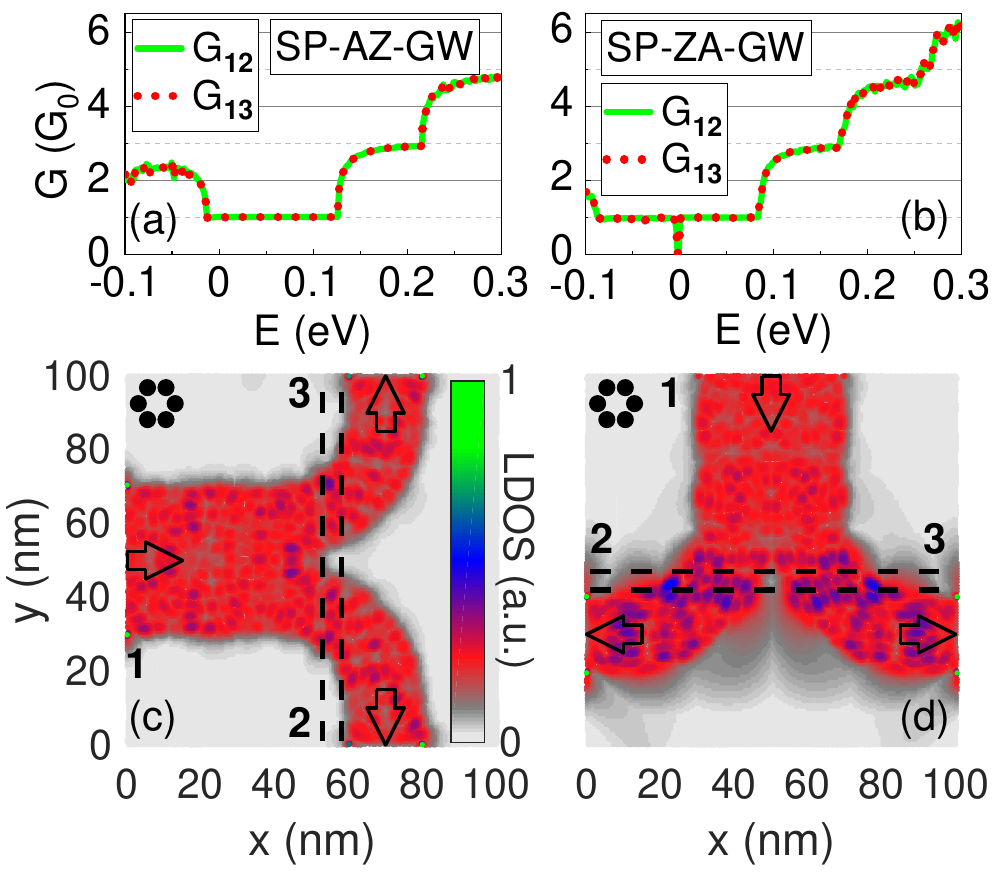}
\caption{(Color online) (a) and (b) \proofreadtrue\changemarker{show the conductance of SP-AZ-GW and SP-ZA-GW with} W$_G$~=~40~nm for stem and W$_G$~=~20~nm for branches. LDOSs of SP-AZ-GW and SP-ZA-GW are presented in (c) and (d) at E~=~0.05~eV.}\label{fig9}
\end{figure}
\proofreadtrue\changemarker{We also considered two  configurations for the split waveguide, labeled by SP-AZ-GW and SP-ZA-GW, in which SP-AZ-GW (SP-ZA-GW) refers to a split waveguide where the orientation of stem is armchair (zigzag), while that of the branches is zigzag (armchair).}
\proofreadtrue\changemarker{Like the case of the L-shape waveguide, drain leads at the end of branches are connected to different interfaces, which are opposite to the interface between source lead and the stem, due to the $90^\circ$ bending of each L-shape waveguide.}
\proofreadtrue\changemarker{The calculated conductance  through different paths (G12 and G13) is shown in Figs.~\ref{fig9}(a) and \ref{fig9}(b), in which the first subindex (i.e., 1) refer to the stem while the second subindex (i.e., 2 or 3) refers to each branch.}
Conductance for both paths in the three-terminal SP-AZ-GW show similar trend to that of the 20~nm straight ZO-GW. 

\begin{figure}
\includegraphics[width=8.2cm]{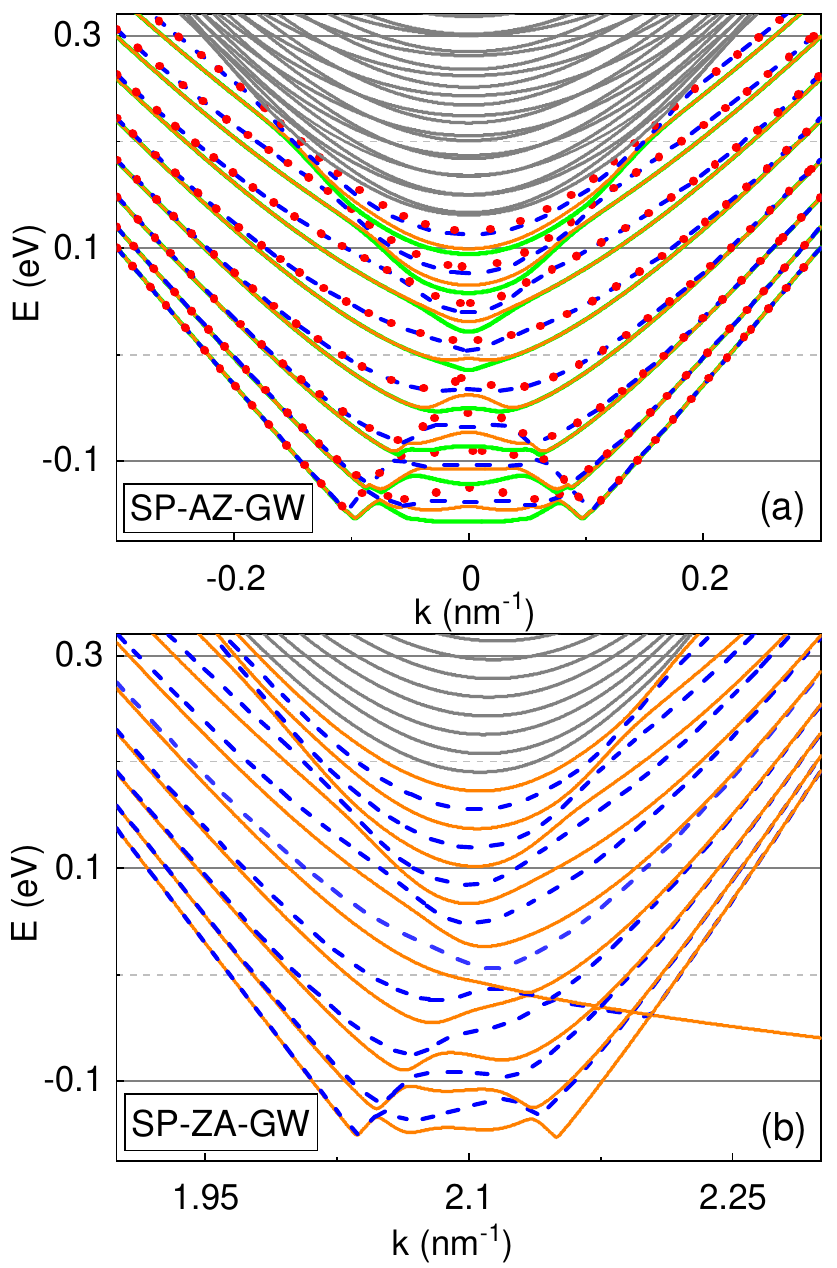}
	\caption{\proofreadtrue\changemarker{Energy band structures for the split waveguides around the splitting area. (a) Band structure calculated for the supercell indicated by the dashed rectangle (an armchair supercell) in Fig.~\ref{fig9}(c). (b) Band structure calculated for the supercell indicated by the dashed rectangle (a zigzag supercell) in Fig.~\ref{fig9}(d). Different color lines (except for the gray lines) are used to distinguish the four-fold and two-fold degenerated subbands in (a) and in (b), respectively. Upper solid gray lines denote the bands corresponding to the wavefunctions that are not confined in the waveguide.}}
\label{fig10}
\end{figure}

The conductance of SP-ZA-GW also follows a similar pattern to the 20~nm \proofreadtrue\changemarker{straight} AO-GW, which can be recognized by \proofreadtrue\changemarker{the small dip in} conductance around E~=~0~eV (see Fig.~\ref{fig9}(b)).  
\proofreadtrue\changemarker{Together with the small dip observed in other armchair drain-based waveguides, we concluded that the nature of drain leads (metallic or nonmetallic, and width) significantly determines the conductance profile of various types of graphene waveguides, regardless of their bending geometries~\cite{mosallanejad2018coherent}.}
\proofreadtrue\changemarker{Again, we plotted the normalized LDOS of split waveguides for each configuration in Figs.~\ref{fig9}(c) and \ref{fig9}(d) to depict the confinement at E~=~0.05~eV corresponding to the first conductance plateau.}
In addition, quasi-one dimensional band structures for selected supercells around the splitting point, indicated by dashed rectangles in Figs.~\ref{fig9}(c) and \ref{fig9}(d), are plotted in Figs.~\ref{fig10}(a) and \ref{fig10}(b), respectively. We have chosen these segments of the scattering area, because they \proofreadtrue\changemarker{give us the information of} the energy bands at the beginning of two independent branches. 
\proofreadtrue\changemarker{The calculated energy} bands show the two-fold (Fig.~\ref{fig10}(b)) and four-fold (Fig.~\ref{fig10}(a)) degeneracy for supercells with zigzag (Fig.~\ref{fig9}(d)) and armchair (Fig.~\ref{fig9}(c)) edges.
\proofreadtrue\changemarker{The number of energy bands in the presence of branches has doubled compared to the band structures of the straight graphene waveguides (see Figs.~\ref{fig5}(b)-\ref{fig5}(d) and Figs.~\ref{fig5}(f)-\ref{fig5}(h)).}
\proofreadtrue\changemarker{Each of the two-fold energy bands in Fig.~\ref{fig10}(b) can be attributed to a non-degenerate energy band belongs to each branches.}
\proofreadtrue\changemarker{Similarly, one can divide the four-fold degenerate energy bands of SP-AZ-GW in Fig.~\ref{fig10}(a) into two two-fold degenerate bands resulting from each branch}. 
Moreover, the symmetry of system along the transport direction \proofreadtrue\changemarker{in the stem part assures} the spatial continuity of energy channels along each branch segments. 
Therefore, the incoming wave has equal probability to scatter into each \proofreadtrue\changemarker{branch at the splitting point and results in ballistic transport from splitting point to drains.}
This justifies the similarity \proofreadtrue\changemarker{of conductance between two branches, as can be observed in G$_{12}$ and G$_{13}$ (see Figs.~\ref{fig9}(a) and \ref{fig9}(b)).}
\proofreadtrue\changemarker{
\subsection{C. Upscaling Graphene Waveguides}
Although the recursive NEGF enable us to perform transport calculations on all the aforementioned examples, the large amounts of memory required by the algorithm renders it incapable of handling structures longer than 200~nm in common computing machines. One solution to this hurdle is to employ a scalable tight-binding approach to examine the quantization of conductance on much larger graphene waveguides. A scalable tight-binding model refers to upscaling the real carbon-carbon bond length (a$_{cc}$) in graphene via a$_{Scale}$~=~S$_f$a$_{cc}$, with the scaling factor S$_f$$>$1~\cite{liu2015scalable}. On the other hand, the nearest hopping energy t$_0$ must be modified to t$_0$/S$_f$ to keep the energy band structure unchanged in the low energy regime. First, we performed the transport study on 20~nm waveguides (i.e., our early example with armchair and zigzag edge orientations) with two different scaling factors 2 and 4. Note that the size of waveguide is fixed, so the increase of the scaling factor actually reduces the number of carbon atoms in the calculation. Conductance of the scaled graphene waveguides with both orientations along with conductance of the non-scaled devices (S$_f$~=~1 as a reference) have been shown in Figs.~\ref{fig11}(a)-\ref{fig11}(b).  
\begin{figure}
	\includegraphics[width=8.6cm]{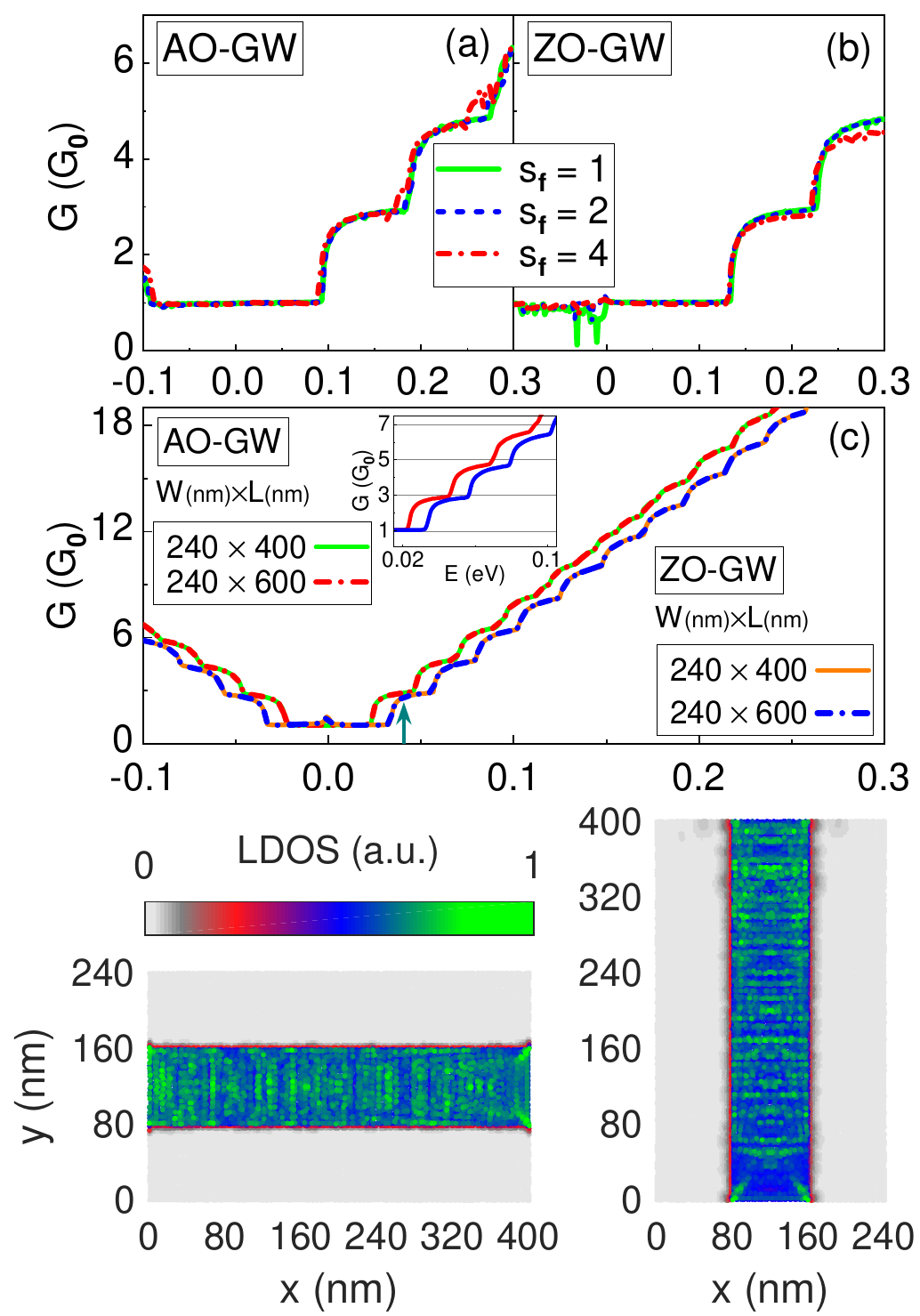}
	\caption{(Color online) \proofreadtrue\changemarker{(a) and (b) show the conductance of a 20~nm wide AO-GW and ZO-GW calculated by the scalable tight-binding model with a scaling factor S$_f$~=~1,~2, and 4, respectively. Note that S$_f$~=~1 corresponds to the original tight binding model (green-solid line). (c) Conductance calculated by the scalable tight-binding model with S$_f$~=~4 for longer graphene waveguides in both orientations. (d) and (e) LDOSs of AO-GW and ZO-GW. LDOSs are extracted at the energy value that is indicated by an arrow in (c).}}\label{fig11}
	\end{figure} 
Conductance calculated by the scalable model shows reasonable consistency with that calculated using the real model. However, we detected two minor differences. First, the resulting conductance of the scaled model in the case of AO-GW delivered noisier conductance in the upper range of Fermi energy. Secondly, the conductance of the scalable model with a larger scaling factor tended to lower the original spacing between plateaus in the case of ZO-GW. Furthermore, we performed a transport study for 80~nm waveguides with L~=~400~nm and L~=~600~nm in both orientations using the scaling factor S$_f$~=~4. The results are plotted in Fig. 11(c). In general, conductance in both types of large-scale graphene waveguides showed reduced spacing between plateaus (less than 2G$_0$~=~$4e^2/h$) and became more fractional with respect to nG$_0$ (n~=~1,~3,~5,\ldots see inset in Fig.~\ref{fig11}(c)). Spacing between plateaus in zigzag oriented waveguides is more uniform than in armchair oriented waveguides, which has presented a series of hardly distinguishable plateaus for E$>$0.1~eV. Conductance of the longer devices (L~=~600~nm) are similar to results produced with the L~=~400~nm devices in both orientations. These results suggest that the effect of valley degeneracy gradually disappears in a longer waveguide, as indicated by the reduced spacing between plateaus (less than 2G$_0$), when the scaled model is applied. Two examples of normalized LDOS, for 80~nm-wide graphene waveguides in both orientations, are plotted in Figs.~\ref{fig11}(d)-\ref{fig11}(e). These show the effect of confinement achieved by the quantum well in the scalable tight-binding model.
In summary, our results show that a small-width graphene waveguide is capable of delivering quantized conductance with the scalable model as long as the well potential is deep enough, which is in contrast to the shallower quantum wells used in Ref.~\cite{rickhaus2015guiding}.}

\section{IV. Conclusion}  
To conclude, by applying the Non-equilibrium Green’s function, we have investigated the transport property of straight and various bent graphene waveguides with two types of edge orientations, i.e., armchair and zigzag configurations. For the straight waveguides, we have shown that the width of side-barrier has little effect on the conductance, while the nature (metallic or non-metallic) and width of the source/drain leads plays an important role in waveguide conductance profiles. In particular, the conductance of waveguides is found to primarily follow the conductance property of the drain terminal in the case of ZO-GW under the 3NN approximation. The conductance in both armchair and zigzag oriented waveguides can be quantized by steps of $4e^2/h$ in a similar manner, but the zigzag oriented waveguide shows a longer first plateau in cases where its drain terminal possesses zero energy modes. From a series of analyses into conductance characteristics, we have observed that the conductance of bent graphene waveguides is similar to that of their straight counterparts, regardless of the bending degree of the guide region for different geometric configurations. LDOS maps for all configurations have shown a good capacity to confine charged particles at the Fermi energies corresponding to the first few conductance plateaus. Moreover, we have employed the scalable tight-binding model to effectively capture the conductance of large-scale straight graphene waveguides. The conductance profile of large-scale graphene waveguides with both orientations exhibits quantized steps close to $4e^2/h$, while the spacing between plateaus is sensitive to the employed scaling factor. . Altogether, this study has demonstrated that coherent transport can be achieved in various electrically gated graphene waveguides with different edge orientations. The conductance quantization realized in straight and highly bent graphene waveguides is promising for application of graphene in modern nanoelectronic devices and thus making all-graphene integrated circuits possible in the future.             
\section[sec:level1]{Acknowledgments}
This work was financially supported by the National Key Research and Development Program of China (Grant No. 2016YFA0301700), NNSFC (Grant No. 11625419 ), the Strategic Priority Research Program of the CAS (Grant Nos. XDB24030601 and XDB30000000), the Anhui initiative in Quantum information Technologies (Grants No. AHY080000). This was also supported by Chinese Academy of Sciences and The World Academy of Science for the advancement of science in developing countries.
	
\end{document}